\documentclass[square]{ws-procs11x85}
\usepackage{balance} 

\newcommand{\met}{\hbox{E\kern-0.5em\lower-0.1ex\hbox{/}}_T}

\begin{document}

\twocolumn[
\title{Emission and power of blazar jets}

\author{Gabriele Ghisellini}

\address{I.N.A.F. -- Osservatorio Astronomico di Brera
I--23807 Merate (Lecco), Italy\\E-mail: gabriele.ghisellini@brera.inaf.it}


\begin{abstract}
Through the modelling of the Spectral Energy Distribution 
of blazars we can infer the physical parameters required
to originate the flux we see.
Then we can estimate the power of blazar jets in the form of matter and
fields.
These estimate are rather robust for all classes of blazars, although
they are in part dependent of the chosen model (i.e. leptonic rather than
adronic).
The indication is that, in almost all cases, the carried Poynting flux
is not dominant, while protons should carry most of the power.
In emission line blazars the jet has a comparable, and often
larger, power that the luminosity of the accretion disk.
This is even more true for line--less BL Lacs.
If the jet is structured at the sub--pc scale, with a fast spine 
surrounded by a slower
layer, then one component sees the radiation of the other boosted,
and this interplay enhances the Inverse Compton flux of both.
Since the layer emission is less beamed, it can be seen also at large viewing
angles, making radio--galaxies very interesting GLAST candidates.
Such structures need not be stable components, and can form and disappear 
rapidly.
Ultrafast TeV variability is challenging all existing models, 
suggesting that at least parts of the jets are moving with
large bulk Lorentz factors and at extremely small viewing angles.
However, these fast ``bullets" are not necessarily challenging
our main ideas about the energetics and the composition of the bulk 
of the jet.
\end{abstract}
\vskip12pt  
]

\bodymatter

\section{Introduction}

EGRET, onboard {\it CGRO}, and the ground based
Cherenkov telescopes showed that blazars are the most
powerful high energy extragalactic emitters, and allowed
to know the bolometric luminosity of these objects.
Now, just after the launch of {\it AGILE} and just before
{\it GLAST}, we are preparing for (and already tasting) the 
possibility to have simultaneous data in the optical,
X--rays and the GeV bands (and possibly the TeV one).
This will be possible mainly by {\it Swift}: its rapid slew and flexible 
scheduling will ensure good quality optical and X--ray data
while the $\gamma$--ray observations are still ongoing.
What was an exception in the EGRET era, will be routine.
This is therefore a risky and at the same time healthy time
to put forward new ideas concerning blazars, that {\it GLAST} can falsify.
In this contribution, I will try to summarize how the 
power of jets can be derived, and what inferences can we draw
from that.
I will discuss the ultra--fast TeV variability
recently observed in PKS 2155--304, arguing that, contrary
to previous claims, it is unlikely that the jet of this source
is magnetically dominated.
Furthermore, I will point out that the fact that the TeV emitter BL Lac 
objects are also the least powerful blazars opens up the possibility
to slow down their jets by the Compton rocket effect.

\begin{figure}
\vskip -0.2 true cm 
\hskip -0.5 true cm 
\centerline{\psfig{figure=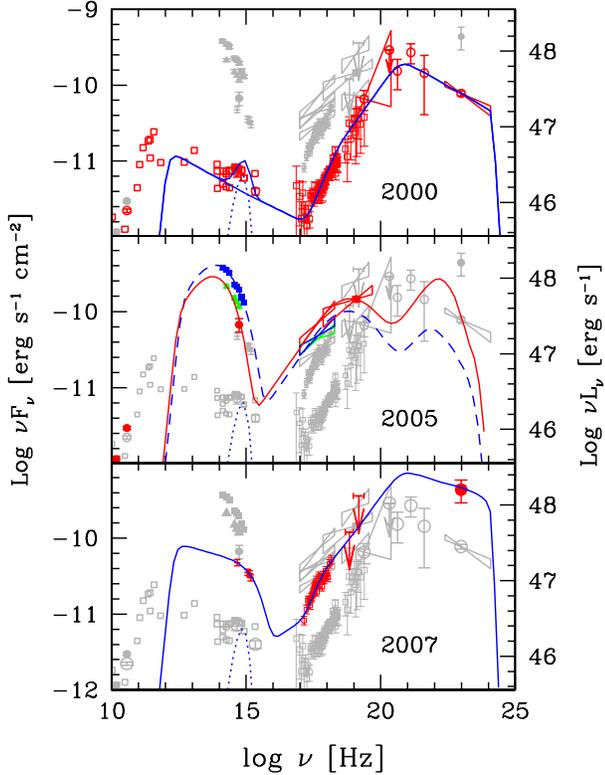,width=11truecm}}
\vskip -0.5 true cm 
\caption{SEDs of 3C 454.3 at different epochs. 
Top panel: the SED in 2000, corresponding to the {\it Beppo}SAX
observations discussed in \cite{tav02}. 
The other points are not simultaneous (see \cite{pian06} 
and references therein).
Mid panel: the SED during the huge optical flare in 2005, as described 
in \cite{pian06, giommi06}.  
Bottom panel: the SED on July 26, 2007, as observed by {\it AGILE}\cite{agile07},
and {\it Swift}. 
The optical flux in the $R$ band comes from the Tuorla observatory.
The optical and X--ray fluxes are corrected for Galactic extinction
($A_V=0.355$).
The solid and dashed lines correspond to our modelling.
The dotted line is the contribution from the accretion disk (assumed
to be a simple black--body). From \cite{gg07}. 
}
\label{454}
\end{figure}

\section{From the SED to the jet power}

Fig. \ref{454} shows three states of 3C 454.3, fitted
with a simple homogeneous, one zone, synchrotron inverse Compton model\cite{gg07}.
The main difference among the three states is the bulk Lorentz
factor $\Gamma$, assumed to be a function of where the jet produces
most of the flux.
This model, proposed and discussed by Katarzynski \& Ghisellini\cite{kris07},
assumes that when the jet dissipates closer in, it has a smaller
$\Gamma$ than when it dissipates at larger distances from the black hole.
This has a great impact on the ratio between the inverse Compton to
synchrotron peak levels $L_{\rm c}/L_{\rm s}$: for small $\Gamma$ 
and more compact dissipation regions, 
it is likely that the magnetic field is larger, and the
external photons, thought to originate in the broad line region (BLR)
are seen less boosted, resulting in a relatively small
$L_{\rm c}/L_{\rm s}$. Viceversa for larger $\Gamma$.
As can be seen, the resulting spectral energy distribution (SED)
can change dramatically in selected frequency ranges, without
the need to a large change of the overall jet power.
This model can be contrasted by the recent idea by Sikora et al. 
\cite{sikora08}, in which the 2007 high $\gamma$--ray state
is attributed to a dissipation region {\it even beyond} the BLR.
At these distances, the magnetic field is small, and the external radiation
is produced by a $\sim$10 pc scale dusty torus.
Variability of the $\gamma$--ray flux is clearly a diagnostic
between the two possibilities.
Fig. \ref{454} shows also that when we see a very hard X--ray spectrum
we have the best chance to measure the low energy cut-off of the
electron distribution, emitting, by the external Compton mechanism, 
in the X--ray range. 
In particular, for 3C 454.3, the electron
distribution is required to extend down to $\gamma_{\rm min}\sim 1$.
This is the energy region where most of the electrons are, and
knowing $\gamma_{\rm min}$ is crucial to determine the power
that the jets carries in the form of bulk motion of particles.

We do not have the same information for BL Lac objects, that can be
fitted by the SSC model without the need of an external component
(but see below from the alternative spine--layer scenario).
On the other hand, for these sources it is less important to know
$\gamma_{\rm min}$, since the average energy of their electrons is higher,
approaching the proton rest mass energy. 
The kinetic power of the jet in this case depends less on the total
number of electrons (and protons) and more on the electron mean energy.

\begin{figure}
\vskip -0.7 true cm
\centerline{\psfig{figure=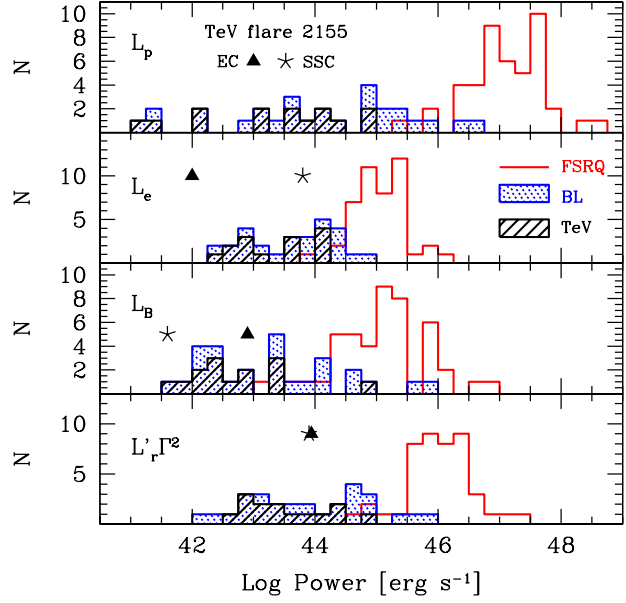,width=9.5truecm}}
\vskip -0.7 true cm
\caption{The power of blazar jets in bulk motion of protons
($L_{\rm p}$, assuming one proton per emitting electrons),  
relativistic electrons ($L_{\rm e}$), magnetic field ($L_{\rm B}$) and
in radiation $L_{\rm r}$.
The star and the triangle correspond to the values of 2155--304
during the TeV flare of 28 July 2006, whose SED has been fitted
with $\Gamma=50$ and with the SSC and EC models (see Fig. \ref{2155sed}).
Adapted from \cite{cg08, gt08}. 
}
\label{istokin}
\end{figure}

\section{Matter dominated jets}

Fig. \ref{istokin} shows the distribution of powers that the jet carries
in the form of bulk motion of protons ($L_{\rm p}$, assuming one proton per electron),
relativistic emitting electrons ($L_{\rm e})$, magnetic field 
($L_B$) and radiation ($L_{\rm r} = \Gamma^2 L^\prime_{\rm r}$,
where $L^\prime_{\rm r}$ is the comoving luminosity).
These powers result from the fitting of a sample of blazars for which
we have high energy observations\cite{cg08}.  
Bear in mind a possible bias: we are selecting
sources in an active state. 
All fits assume a simple one--zone, leptonic, synchrotron 
inverse Compton model.
The different histograms represent FSRQs, ``classical" BL Lacs
and TeV BL Lacs. 
Note that:

\noindent
{\bf The jet power, in FSRQs, is large.}
Estimating the luminosity of the accretion disk from the 
broad emission lines, and sometimes from the directly visible
blue bump emission, we derive that the jet power is often greater than 
the disk luminosity.
This is even more true for BL Lac objects, where no lines are seen.

\noindent
{\bf The power in Poynting flux ($L_B$) is small.}
Even smaller than $L_{\rm r}$. 
This argues against the idea that blazar jets are magnetically dominated 
at all scales (e.g.\cite{bland02, lyutikov03}),
if the magnetic field of the emitting region is representative of
the overall magnetic field at these scales.
A small magnetic field in the emission region  
is expected, since these sources are the brightest
$\gamma$--ray emitters: 
they emit more by the Inverse Compton process than by synchrotron,
and this limits the value of the magnetic field.

\noindent
{\bf TeV BL Lacs have the least powerful jets.}
This suggests that it is easier to make them decelerate
even at the pc scale.

\begin{figure}
\vskip -0.5 true cm
\centerline{\psfig{figure=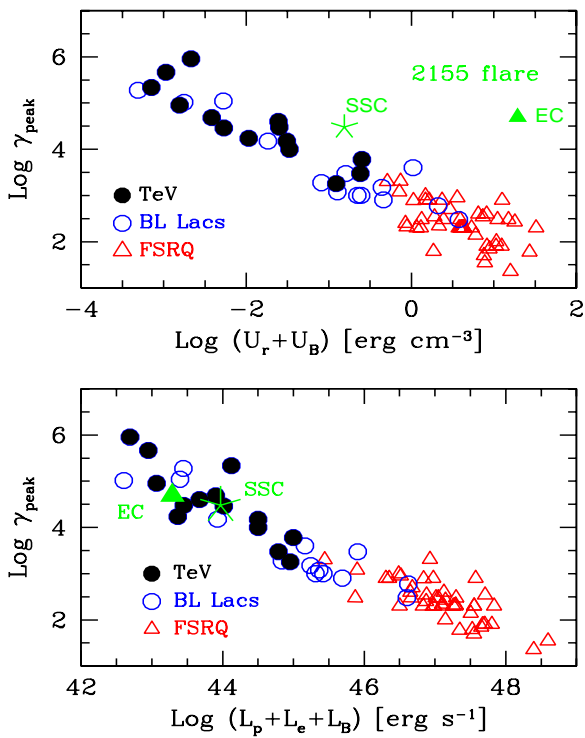,width=12truecm,height=11truecm}}
\vskip -0.5 true cm
\caption{The random Lorentz factor, $\gamma_{\rm peak}$, of the electrons
emitting at the peaks of the SED vs the comoving energy density
(magnetic plus radiative; top panel) and the total jet power (bottom panel).
The star and the triangle correspond to the values of PKS 2155--304
during the TeV flare of 28 July 2006, whose SED has been fitted
with $\Gamma=50$ and with the SSC and EC models (see Fig. \ref{2155sed}).
Adapted from \cite{cg08}.  
}
\label{gpeak}
\end{figure}

\subsection{The blazar sequence}

There are two versions of the so called ``blazar sequence":
the first is purely phenomenological: the observed SEDs
show that the two broad peaks (one in the IR--soft X--rays, the
second in the MeV--GeV and sometimes TeV bands) 
shift to lower frequencies as the observed bolometric 
luminosity is increased. At the same time, the Compton dominance
(the $L_{\rm c}/L_{\rm s}$ ratio) increases \cite{fossati98}, 
see also Maraschi et al. in these proceedings).
The second version of the blazar sequence explains this
phenomenological behavior in terms of increased radiative cooling 
as the total luminosity increases.
Electrons emitting at the peaks of the SEDs have smaller energies
as the luminosity, thus the cooling, increases \cite{gg98, gcc02}.
Fig. \ref{gpeak} shows $\gamma_{\rm peak}$ vs the
comoving energy density (magnetic plus radiative; top panel)
and vs the total jet power (bottom panel).
Different symbols identify FSRQs, ``classical" BL Lacs and TeV BL Lacs.
The general trend is evident: $\gamma_{\rm peak}\propto (U_B+U^\prime_{\rm r})^{-1}$
for large $\gamma_{\rm peak}$, while two branches appear at small
values of $\gamma_{\rm peak}$ [$\propto (U_B+U^\prime_{\rm r})^{-1}$ and
$\propto (U_B+U^\prime_{\rm r})^{-1/2}$].
The bottom panel shows another well defined trend:
$\gamma_{\rm peak}\propto L_{\rm j}^{-3/4}$, where 
$L_{\rm j}=L_{\rm p}+L_{\rm e}+L_B$.
These demonstrates (in the context of the used model) 
a tight link between the amount of radiative cooling
and the power of blazar jets.
It also illustrates that TeV BL Lacs have the least powerful jets 
and the most energetic electrons.
This fact bears an important consequence, as discussed below.

\section{Ultrafast TeV variability}

\begin{figure}
\vskip -0.7 true cm 
\center
\centerline{\psfig{figure=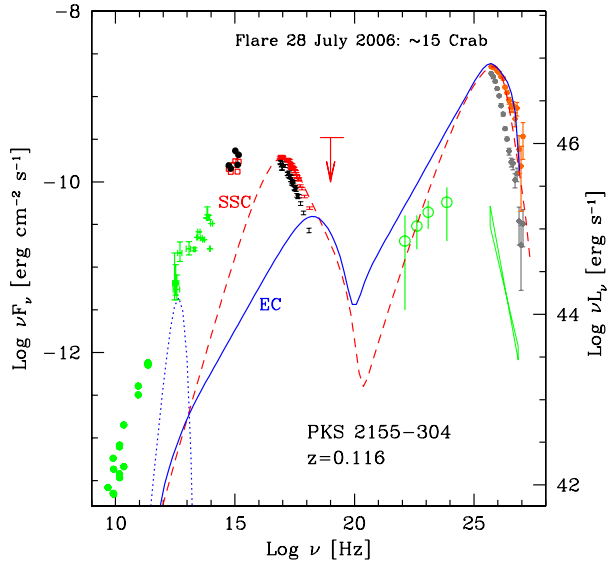,width=9truecm}}
\vskip -0.5 true cm
\caption{
The SED of PKS 2155--304. 
TeV data correspond to the flare of 28 July 2006,
while X--ray and optical data are not strictly simultaneous, but corresponds to 
2 days later (see \cite{foschini07}).
We modelled the SED with an SSC (dashed line) and EC (solid line) 
assuming $\Gamma=50$, a viewing angle of 1$^\circ$ and a size of the emitting region
of $R\sim 3\times 10^{14}$ cm, corresponding to $t_{\rm var}\sim$200 seconds.
The dotted line corresponds to the distribution of seed photons assumed for the EC model.
It has been chosen in order to maximize the high energy output of the EC emission,
without overproducing the far IR observed flux. Adapted from \cite{gt08}.
}
\label{2155sed}
\end{figure}

The increased sensitivity of the new generation of Cherenkov telescopes
allows to observe flux variations down to the minute timescales,
for the brightest sources.
Indeed, PSK 2155--304\cite{aha07} 
and Mkn 501\cite{albert07} 
showed variations on $t_{\rm var}=$3--5 minutes.
Particularly challenging is the case of PKS 2155--304, because the
variations occurred during an overall very active state of the source
with an observed TeV luminosity of $\sim 10^{47}$ erg s$^{-1}$
(see Fig. \ref{2155sed}).
The usual way to infer the size from the variability timescale
is $R<ct_{\rm var}\delta$ where $\delta$ is the Doppler factor.
Begelman, Fabian \& Rees \cite{begelman08} 
pointed out that one consequence of such a small 
$t_{\rm var}$ is that these timescales are no longer indicative of 
the size of the black hole (as instead are in the 
``internal shock scenario"\cite{sikora94, gg99}).
The other consequence is that to avoid to have a too compact source, with
the accompanying problem of $\gamma$--ray absorption through the 
$\gamma$--$\gamma$ $\to e^\pm$ process\cite{dondi95}, 
one is obliged to increase the bulk Lorentz factor of the emitting
region (hence $\delta$) to values close to 50 or more 
(close to the ones of Gamma Ray Bursts). 
In turn, such large $\Gamma$ makes any external photon source strongly 
boosted in the comoving frame, favoring the EC process.
At the same time, the viewing angle, to have $\delta\sim\Gamma$, must be of the 
order of $1^\circ$ or less. 
Therefore Begelman et al.\cite{begelman08} suggest that the EC process is the main 
radiation mechanism.
They also argue that, since to produce TeV photons one needs
highly energetic electrons, the jet could be particle starved (i.e. one needs
fewer electrons to produce the radiation we see, if they are at high energies).
Therefore the jet should be magnetically dominated.

Fig. \ref{2155sed} (see \cite{gt08}) 
shows the TeV data corresponding to the flare (observed and
de--absorbed, according to a low--medium value of the IR background),
together to X--ray and optical data taken a few days later\cite{foschini07}. 
The shown models are a pure SSC model, with no external photons,
and an EC model with external photons produced very close to the
emitting region, at frequencies compatible with scatterings in the Thomson
regime, but whose flux does not overproduce the observed far IR data of the source.
In other words, we have maximized and optimized the possible contribution of the 
external component, imposing, at the same time, the maximum possible magnetic field,
to see if the jet of PKS 2155--304 can be magnetically dominated.
For both models, $\Gamma=50$, the viewing angle is 1$^\circ$ and then $\delta=56$.
The size of the emitting region is $R\sim 3\times 10^{14}$ cm, yielding
$t_{\rm var}\sim 200$ s.
Both the SSC and the EC model can reproduce the
high energy data, but the EC model cannot contribute much to the X--ray flux,
which can instead be reproduced by the SSC model.
In Fig. \ref{istokin} we show the jet powers of PKS 2155--304
according to the two models.
Note that $L_{\rm B}$ is not dominant even in the EC model, where
the enhanced radiation energy density (due to the strong boosting)
allows a larger magnetic field with respect to the SSC model ($B=3$ G vs 0.58 G,
respectively).
Note also that the power corresponding to the produced radiation ($L_{\rm r}$)
barely corresponds to $L_{\rm p}+L_{\rm e}$  for the SSC case, and is larger in
the EC model. 
In Fig. \ref{gpeak} we show the value $\gamma_{\rm peak}$ of PKS 2155--304
vs the comoving energy densities and jet power.
The conclusion is that although the EC model allows (and requires) a larger magnetic
field, the jet remains matter dominated even with $\Gamma=50$.
At the same time, the EC model gives a worse fit (we cannot account for the X--ray flux)
with respect to the SSC model, although a variant of the EC model, in which the variable
TeV emission is produced by a very fast ``needle" immersed in a larger and slower
``normal" jet, gives the best results \cite{gt08}.
Finally, the SSC model gives values of $\gamma_{\rm peak}$ and $U^\prime$ 
in agreement with the general trend observed for other TeV blazars.\cite{cg08}

\section{Structured and decelerating low power jets}

There are several lines of evidence that the jets in low power FR I
radio--galaxies, hence the jet of BL Lac objects, can have
a fast inner spine surrounded by a slower layer.
The evidences come from statistical arguments (unifying FR I with 
BL Lacs\cite{chiab00}), 
and direct radio imaging\cite{giroletti04, swain98, owen89, giovannini99}.
Furthermore, at the VLBI scale, the jets of low power TeV BL Lacs
are slow, often subluminal\cite{ep02, pe04}.
This is in marked contrast with more powerful jets.
A slow motion at the VLBI scale is also at odd with with the need 
of bulk Lorentz factors exceeding 20 for the sub--pc 
scale of TeV BL Lacs.
At these scales, where most of the emission is produced, 
a large $\Gamma$ is required not only to avoid the $\gamma$--$\gamma \to e^\pm$ 
process\cite{dondi95, begelman08}, but also to account for the
large frequency separation of the two peaks of the SED\cite{tav98}.

To explain the latter properties,
Georganopoulos \& Kazanas \cite{georga03} have postulated
that the jet experiences a strong deceleration at the sub--pc scale.
The jet continues to emit during the deceleration phase,
and the emission at different distances from the black hole
is beamed differently. 
The region at the start of the deceleration region sees the radiation 
produced farther away along the jet beamed in its direction. 
Viceversa, the regions moving at lower speeds
see the radiation of the inner jet more beamed.
In this way there is an enhancement of the radiation energy density
seen in the comoving frames
with a consequent enhancement of the inverse Compton emission.

Ghisellini, Tavecchio \& Chiaberge \cite{gtc05}, following this idea
and the above mentioned evidences, proposed that the jet can be 
structured, with a cospatial fast spine surrounded by a slow layer.
Both components emit, and the radiation produced by one is seen
boosted by the other, enhancing the inverse Compton emission of both
components.

This structure influences the jet propagation.
Since the typical energies of the emitting electrons in low power jets 
are very large (see Fig. \ref{gpeak}), the accompanying protons are 
not important for the jet dynamics.
Then if the jet has i) a small power and ii) large electron energies,
it can decelerate by the Compton rocket effect.
In fact, the electrons in the spine efficiently scatter the radiation
produced by the layer. 
In the spine comoving frame, this emission {\it is not isotropic} 
(contrary to the synchrotron and SSC emission): if the produced radiation
is a good fraction of the electron energy (as it seems, see Fig. \ref{istokin}),
then the jet decelerates\cite{gtc05}.

The layer, having a smaller bulk Lorentz factor ($\sim$a few), 
emits in a cone much wider than the spine.
Observers at large viewing angles see the layer, not the spine.
Since also the layer copiously emits at high energies (by scattering
spine photons), this idea predicts that several radio--galaxies
should be detected by {\it GLAST}\cite{gtc05}.
In one case, M87, we might have already seen the TeV
emission of the layer, as discussed below.

\begin{figure}
\vskip -0.9 true cm
\center
\centerline{\psfig{figure=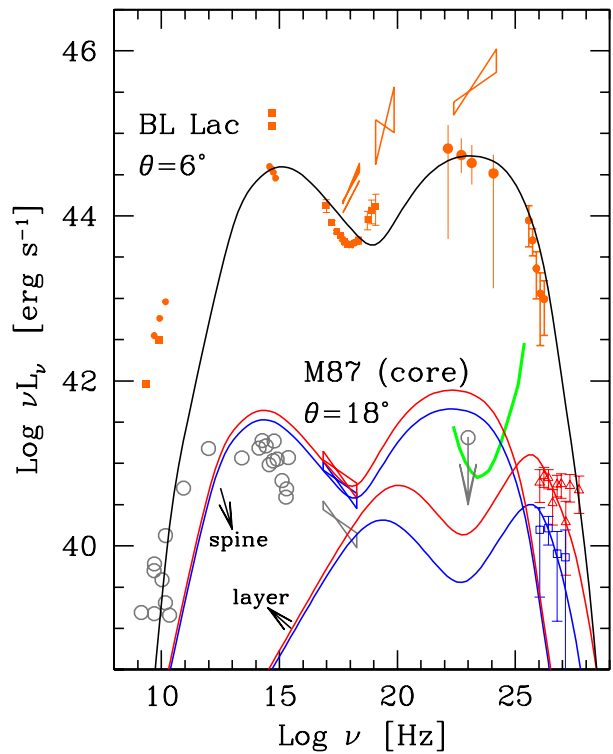,width=12truecm}}
\vskip -0.7 true cm
\caption{
SED of the core of M87 together with the
H.E.S.S. spectra taken in 2004 and 2005\cite{aha06}. 
The upper model is the emission from the spine that would 
be observed at 6$^\circ$ from the jet axis, compared with the
SED of BL Lac. From \cite{tg08}.
}
\label{m87}
\end{figure}

\subsection{TeV radiogalaxies: the case of M87}

M87 is the only non--blazar extragalactic TeV source\cite{aha06}
among the 20 already detected (see 
www. mppmu.mpg.de/$\sim$rwagner/sources).
Even if it is not possible to resolve the TeV emission region, 
the TeV flux variability ($\sim$ days) suggests a compact source.
Among the sites proposed as emission regions of the TeV
flux are the resolved jet (in particular the so--called knot HST--1)
and the unresolved base of the jet, in analogy with blazars.

HST--1 is located at 60 pc (projected) from the core of of M87. 
It showed spectacular activity in the past, and
this extreme phenomenology is described by \cite{stawarz06} 
assuming that HST--1 marks the recollimation shock of the jet. 
As such, HST--1 is thought to be a
rather efficient particle accelerator and thus a possible source of
intense TeV radiation, but its location at large distances from the core
contrasts with the rapid TeV variability, that could instead be 
easily reproduced if the TeV emission is produced in the
more compact ``blazar"--like region.

The determination of the TeV emission site is rather important:
if the emission site will be eventually identified as knot HST--1, 
this will have a broad impact on the current view of relativistic jets. 

We\cite{tg08} have then tried to model the SED of M87 with a spine--layer jet,
assuming that the TeV flux is produced by the layer, while the rest
of the spectrum is produced by the spine, dominating
the emission at lower energies even at the relatively large
M87 viewing angle ($\sim$$18^\circ$, see Fig. \ref{m87}).
The emission region is at the same distances than in blazars.
``De--coupling" the two peaks of the SED (one is made by the spine,
the other by the layer) solves the most severe difficulty faced by standard
one--zone SSC models when applied to the SED of M87.
The price to pay is to increase (to double) the number of parameters.
For this reason we not only fitted the M87 SED, but were careful to see
the model predictions for observers located at smaller angles.
As can be seen in Fig. \ref{m87}, the blazar ``paired" to M87 
(for a viewing angle of 6$^\circ$) has a SED
closely resembling the one of BL Lac itself.
This is only a ``consistency" check, but it is encouraging.

\section*{Acknowledgments}
I gratefully thank Annalisa Celotti, Luigi Foschini and Fabrizio Tavecchio.

\end{document}